\documentclass[conference]{IEEEtran}

\renewcommand{\columnsep}{5mm}
\renewcommand{\textwidth}{6.8in}

\ifCLASSINFOpdf

\else

\fi

\usepackage{graphicx}
\usepackage{amsmath}
\usepackage{amssymb}
\usepackage{tabularx}
\usepackage[english]{babel}
\usepackage{multirow}

\usepackage{algorithm,algpseudocode}
\algnewcommand{\Inputs}[1]{%
  \State \textbf{Inputs:}
  \Statex \hspace*{\algorithmicindent}\parbox[t]{.8\linewidth}{\raggedright #1}
}
\algnewcommand{\Initialize}[1]{%
  \State \textbf{Initialize:}
  \Statex \hspace*{\algorithmicindent}\parbox[t]{.8\linewidth}{\raggedright #1}
}
\algnewcommand{\Outputs}[1]{%
  \State \textbf{Outputs:}
  \Statex \hspace*{\algorithmicindent}\parbox[t]{.8\linewidth}{\raggedright #1}
}

\begin{document}

\title{\large{\bf{Identification of relevant diffusion MRI metrics impacting cognitive functions using a novel feature selection method}}}

\author{

\IEEEauthorblockN{\textit{Tongda Xu\textsuperscript{1}, Xiyan Cai\textsuperscript{2}, Yao Wang\textsuperscript{1}, Xiuyuan  Wang\textsuperscript{3}, Sohae Chung\textsuperscript{3}, Els Fieremans\textsuperscript{3},}} 

\IEEEauthorblockN{\textit{Joseph Rath\textsuperscript{4}, Steven Flanagan\textsuperscript{4}, Yvonne W Lui\textsuperscript{3}}}

\IEEEauthorblockA{1. Electrical and Computer Engineering Department, New York University}
\IEEEauthorblockA{2. New York University Shanghai}
\IEEEauthorblockA{3. Department of Radiology, New York University}
\IEEEauthorblockA{4. Department of Rehabilitation Medicine, New York University}
\{x.tongda, xc984, yw523\}@nyu.edu \\
\{Xiuyuan.Wang, Sohae.Chung, Els.Fieremans, Joseph.Rath, Steven.flanagan, Yvonne.Lui\}@nyulangone.org}

\maketitle

\begin{abstract}Mild Traumatic Brain Injury (mTBI) is a significant public health problem. The most troubling symptoms after mTBI are cognitive complaints. Studies show measurable differences between patients with mTBI and healthy controls with respect to tissue microstructure using diffusion MRI. However, it remains unclear which diffusion measures are the most informative with regard to cognitive functions in both the healthy state as well as after injury. In this study, we use diffusion MRI to formulate a predictive model for performance on working memory based on the most relevant MRI features. As exhaustive search is impractical, the key challenge is to identify relevant features over a large feature space with high accuracy within reasonable time-frame. To tackle this challenge, we propose a novel improvement of the best first search approach with crossover operators inspired by genetic algorithm. Compared against other heuristic feature selection algorithms, the proposed method achieves significantly more accurate predictions and yields clinically interpretable selected features (improvement of $r^2$ in 8 of 9 cohorts and up to 0.08). 

\end{abstract}

\IEEEpeerreviewmaketitle

\section{Introduction}

Mild traumatic brain injury (mTBI) is a significant public health issue with millions of civilian, military, and sport-related injuries occurring every year \cite{faul2010traumatic}. Moreover, $20-30\%$ of patients with mTBI develop persistent symptoms months to years after initial injury \cite{voormolen2018divergent}. Cognitive complaints are important due to their significant impact on the quality of life. In this study, we examine the specific cognitive subdomain of working memory in relation to the underlying tissue microstructure by accessing diffusion MRI and predict performance on working memory. Defining specific imaging biomarkers related to cognitive dysfunction after mTBI would not only shed light on the underlying pathophysiology of injury leading to cognitive impairments, but also help to triage patients and offer a quantitative means to track recovery in the cognitive domain as well as track efficacy of targeted cognitive therapeutic strategies \cite{grossman2010mild}. Tools to detect injury, predictive of symptoms are badly needed.

Diffusion MRI is a powerful non-invasive method to probe brain tissue microstructure after mTBI \cite{shenton2012review}\cite{chung2018white}. Diffusion tensor imaging (DTI) and diffusion kurtosis imaging (DKI) have been used to reveal areas of abnormal fractional anisotropy (FA) and mean kurtosis (MK) \cite{shenton2012review} (See Tab.~\ref{tab:metric}). More recently, multi-shell diffusion imaging was used to acquire compartment-specific white matter tract integrity metrics to investigate the biophysical changes in mTBI \cite{chung2018white}. In particular, measures of axon injury in mTBI may be associated with alterations in working memory performance \cite{inpresspaper}\cite{miles2008short}.

A few previous works apply feature analysis to identify injury in mTBI and to predict clinical status of mTBI patients. Lui et al. used Minimum Redundancy and Maximum Relevance (mRMR) approach to identify  the most relevant features for classifying patients between mTBI versus control \cite{lui2014classification}. Minaee et al. proposed a combination of linear regression and exhaustive MRI feature selection to predict neuropsychological (NP)  test scores \cite{RN1}. Though they reported achieving reasonable accuracy, these methods were developed using very small datasets ($<50$ subjects) and explored only a small set of handcrafted imaging features (10-15 features). Due to limited datasets, it is not feasible to apply deep learning to entire brain volumes obtained with multiple diffusion metrics for either task (mTBI classification or prediction of NP scores). To overcome this challenge, Minaee et al. \cite{RN2} applied Adversarial Auto-encoder \cite{makhzani2015adversarial} to extract latent features that could then be used to reconstruct image patches, and adopted a bag of visual words (BoW) representation to describe each metric in each brain region, where the visual words were obtained by clustering the latent features. Despite a high classification accuracy \cite{RN2}, feature selection after BoW was accomplished by greedy forward search, which may produce suboptimal feature subset that is quite far from the optimal one.

There are several other works that use imaging features to study mTBI, such as dictionary learning \cite{kao2016unsupervised} for dimensionality reduction and network-based statistics analysis \cite{mitra2016statistical}. However, since feature selection over a large feature space is prohibitively expensive, these works either 1) are limited in the number of initial features considered, which relies on prior knowledge to handcraft features and may potentially miss the most relevant ones or 2) project an originally large feature dimension to a low dimension space; a downside of such approaches is that the transformed features are often hard to interpret. 

To overcome these limitations, we leverage a powerful feature selection method known as greedy best first search (Greedy BFS) \cite{RN13}, which has been shown to be more effective than the more typically adopted greedy forward or backward search method or the genetic algorithm. We further propose a novel improvement over the Greedy BFS method. First, sufficiently large (280) number of statistic features are extracted from 7 anatomic white matter brain regions and 8 diffusion MRI metrics. Gradient Boosting Tree (GBT) is selected for accurate regression and repeated stratified cross-validation is used to avoid over-fitting. During the search, each feature subset is evaluated by the cross validation $r^2$ score by the GBT model. The proposed improvement to the Greedy BFS method, known as BFS with crossover, uses crossovers to jump over the feature subset graph so that a broader feature subset can be visited to produce a more accurate result.

Compared to using greedy forward or backward search or genetic search, Greedy BFS method yielded greater accuracy in the prediction of working memory subtests'scores from difussion MRI features. The BFS with crossover further improved the accuracy over greedy BFS consistently. Interestingly, the features that were chosen frequently by the BFS with crossover method are those diffusion MRI metrics that represent the underlying tissue microstructure.
\section{Method}
\subsection{Dataset and Feature Extraction}

The dataset contains 154 subjects, among which 70 are normal controls (NC) and 84 are mTBI. Age-appropriate WAIS-IV subtests \cite{sattler2009assessment} were performed to assess performance of working memory, including Digit Span Forward (DSF), Digit Span Backward (DSB), and Letter-Number Sequencing (LNS). For each subset, separate models are developed for the control and mTBI populations, respectively, in order to discover normal and pathologic microstructure features that inform on the working memory. In addition, a combined model is also developed.

Based on previous diffusion studies in mTBI patients, 8 metrics from DTI, DKI and compartment specific white matter modeling \cite{miles2008short}\cite{chung2018white} were chosen, summarized in Table~\ref{tab:metric}. For compartment specific metrics, voxels with FA $<0.4$ were excluded as recommended to interrogate single-fiber orientations \cite{fieremans2011white} \cite{jensen2017evaluating}. Instead of considering the entire brain volume, we compute several statistics of each metric over 7 major white matter brain regions: left rostral (LR), right rostral (RR), left middle (LM), right middle (RM), left caudal (LC), right caudal (RC), and corpus collasum (CC). 5 statistics are computed for each metric and each region: mean, standard deviation, skewness, kurtosis, entropy. In total, there are 280 initial features.

\begin{table}[h!]
\caption{MRI metrics description}
\label{tab:metric}
\centering
\begin{tabular}{|c|c|c|}
\hline
\multicolumn{2}{|c|}{Diffusion Imaging Metrics} & Description                     \\ \hline
\multirow{2}{*}{DTI}          & FA              & Fractional Anisotropy           \\ \cline{2-3} 
                              & MD              & Mean Diffusion                  \\ \hline
DKI                           & MK/AK           & Mean/Axial Kurtosis             \\ \hline
\multirow{4}{*}{Compartment Specific}         & AWF             & Axonal Water Fraction           \\ \cline{2-3} 
                              & DA              & Intra-axonal diffusivity        \\ \cline{2-3} 
                              & De-par          & Extra-axonal axial diffusivity  \\ \cline{2-3} 
                              & De-perp         & Extra-axonal radial diffusivity \\ \hline
\end{tabular}\end{table}

\subsection{Wrapper Feature Selection as Graph Search Problem}

\begin{figure}[!t]
\centering
\includegraphics[width = \linewidth]{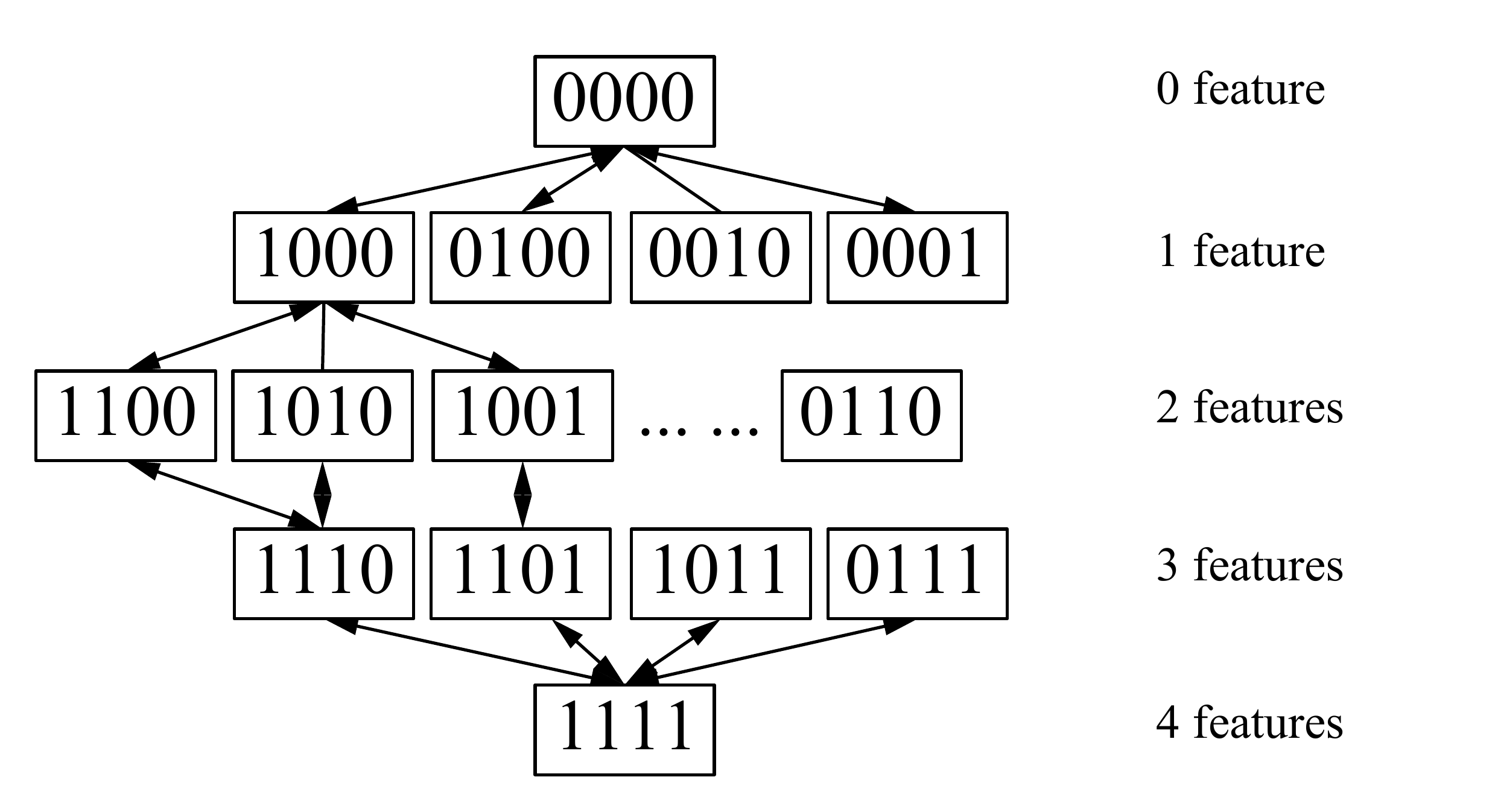}
\caption{An example of 4 feature graph with crossover operator, each node represent a possible feature subset}
\label{graph1}
\end{figure}

There are three main categories of feature selection methods: filter, wrapper and embedded \cite{RN15}. Filter based feature selection ranks the feature subsets based on some criteria such as the correlation between the individual features and the target outcome and the correlations among the features, independent of the prediction/ classification method. Wrapper based approach would train multiple prediction/ classification models using different feature subsets and use validation scores to select the best feature subset. Embedded approach constrains the model parameters related to the input features to be sparse, and conducts feature selection during model construction. In general, filter approach is computationally fastest but often yields sub-optimal feature subsets; whereas the wrapper method is the most accurate but is computationally costly. In this work we follow the wrapper approach.

The wrapper based feature subset selection can be generalized as a graph search problem \cite{RN18}. Consider a dataset with N samples $\{X^{(i)}, y^{(i)}\}_{i = 1,2...N}$, where $X^{(i)}$ represents the features for the $i^{th}$ sample, and $y^{(i)}$ the ground truth outcome. If each data sample $X^{(i)}$ has $M$ features, the number of total possible feature subsets is $2^{M}$.

Then consider a directed weighted graph $G(V, E)$. Each vertex is represented by a binary vector in $M$ dimensions, $V_i = \{0,1\}^{M}$, indicating whether each feature is selected. Two vertices $V_i$ and $V_j$ are connected if there is only one bit difference, which means only the state of one feature is different (See Fig.~\ref{graph1} ). $G(V, E)$ contains $2^M$ vertices, with in-degree and out-degree of each vertex both equals to $M$. The weight of an edge is assigned to be the difference between the performance scores of connected vertices, which is usually calculated through cross-validation \cite{RN18},
\begin{equation}\label{gedge}
E(V_i, V_j) = score(V_j) - score(V_i)
\end{equation}
with
\begin{equation}\label{gedgest}
|V_i - V_j| = 1
\end{equation}
Any path $P$ connecting vertex $V_i$ and another vertex $V_j$ has length equals to the sum of the edge weights along this path:
\begin{equation}\label{gpath}
P(V_i, V_j) = E(V_i, V_{i+1}) + ... +E(V_{j-1}, V_j)
\end{equation}
From the definition of Eq.~(\ref{gedge}), it is easy to show that:
\begin{equation}\label{gfinal}
P(V_i, V_j) = score(V_j) - score(V_i)
\end{equation}
Therefore, the feature selection problem is to find the longest path $P^*$ from vertex $0$ to any possible vertex in graph, which is equivalent to searching the vertex $V^*$ with the best score:
\begin{equation}\label{gbesteq}
V^* \gets \underset{V_i \in G} {argmax} \ score(V_i) - score(0^T)
\end{equation}

\begin{algorithm}[t]
\caption{Greedy Best First Search}\label{alg:astar}
\begin{algorithmic}[t]
\Procedure{Greedy-BFS}{$X,y,patience$}
\State $open \gets \{0^T\}, close \gets \emptyset, best \gets \{0^T\}, i \gets 0$
\While{$True$}
\State $current \gets open.popmax()$ \Comment{pop the node in the open list with maximum score}
\State $close.enqueue(current)$
\If {$current.score > best.score$}
\State $best \gets current$
\State $i \gets 0$
\Else
\State $i \gets i+1$
\EndIf
\If {$i > patience$}
\State \Return $best$
\EndIf
\For {$child \in current.adjacencylist$} 
\If {$child \notin open \bigwedge child \notin close$}
\State $child.crossvalidate(X,y)$
\State $open.enqueue(child)$ \Comment{add child into the open list}
\EndIf
\EndFor
\EndWhile
\EndProcedure
\end{algorithmic}
\end{algorithm}
\begin{algorithm}
\caption{Greedy Best First Search Crossover Operator}\label{alg:astarcomp}
\begin{algorithmic}[t]
\Procedure{Greedy-BFS-X}{$X,y,patience$}
\State $open \gets \{0^T\}, close \gets \emptyset, best \gets \{0^T\}, i \gets 0$
\While{$True$}
\If {$cross \& cross.score>current.score$}
\State $current \gets cross$
\State $open.dequeue(cross)$
\Else
\State $current \gets open.popmax()$ 
\EndIf
\State $close.enqueue(current)$
\If {$current.score > best.score$}
\State $best \gets current$
\State $i \gets 0$
\Else
\State $i \gets i+1$
\EndIf
\If {$i > patience$}
\State \Return $best$
\EndIf
\State $local \gets \emptyset$
\For {$child \in current.adjacencylist$} 
\If {$child \notin open \bigwedge child \notin close$}
\State $child.crossvalidate(X,y)$
\State $local.enqueue(child)$ 
\EndIf
\EndFor
\State $open.merge(local)$ \Comment{merge local queue with open queue}
\State $first \gets local.popmax()$
\State $second \gets local.popmax()$
\State $cross \gets first + second - current$
\If {$cross \notin open \bigwedge child \notin close$}
\State $open.enqueue(cross)$
\EndIf
\EndWhile
\EndProcedure
\end{algorithmic}
\end{algorithm}

\subsection{Greedy Best First Search Algorithm}
Since the graph has $2^M$ nodes, an exhaustive traverse would be prohibitive if $M$ is large. Thus, a heuristic is usually used to avoid exhaustive search without losing accuracy significantly. Some classical heuristic approaches are: sequential feature selection (SFS), Hill Climbing. Meta-heuristic is another family of algorithms that simulates natural phenomena, including simulated annealing (SA), swarm algorithm such as whale optimization (WO) and genetic algorithm (GA). 

In this paper, we revisit and improve a heuristic approach: greedy best first search (Greedy BFS). Greedy BFS is initially proposed for robots' path finding and later applied to feature selection \cite{doran1966experiments} \cite{RN13}. However, this method did not get much traction due to the limited feature subset size and computation power at that time. Recently researchers start to rediscover it and its variations for problems such as sparse representation \cite{RN22}.

As shown in Algorithm~\ref{alg:astar}, greedy BFS algorithm starts at one node and iteratively selects next node $V_i$ maximizing $score(V_i)$. Each time the node with the best score ``current'' node in Algorithm~\ref{alg:astar}) in the priority queue (``open'' queue in Algorithm~\ref{alg:astar}) is popped out, its undiscovered children are evaluated and pushed into the priority queue. This process is repeated until the queue is empty or the best accuracy has not been updated for a certain number of iteration (patience).

Greedy BFS is a superset of sequential floating feature selection (SFFS) \cite{pudil1994floating}, which is in turn a superset of sequential feature selection (SFS). SFS includes greedy forward and backward. When the patience is set to infinity it is equivalent to exhaustive search.

\subsection{Best First Search with Crossover Operator}

Despite its potential, Greedy BFS is not widely applied to feature selection because it is computationally costly. In each step, it evaluates all children of the current vertex, the number of which equals to $M$, the out-degree of vertex. To solve this problem, we propose a novel algorithm combining Greedy BFS search and crossover operator.

\begin{figure}[h]
\centering
\includegraphics[width = \linewidth]{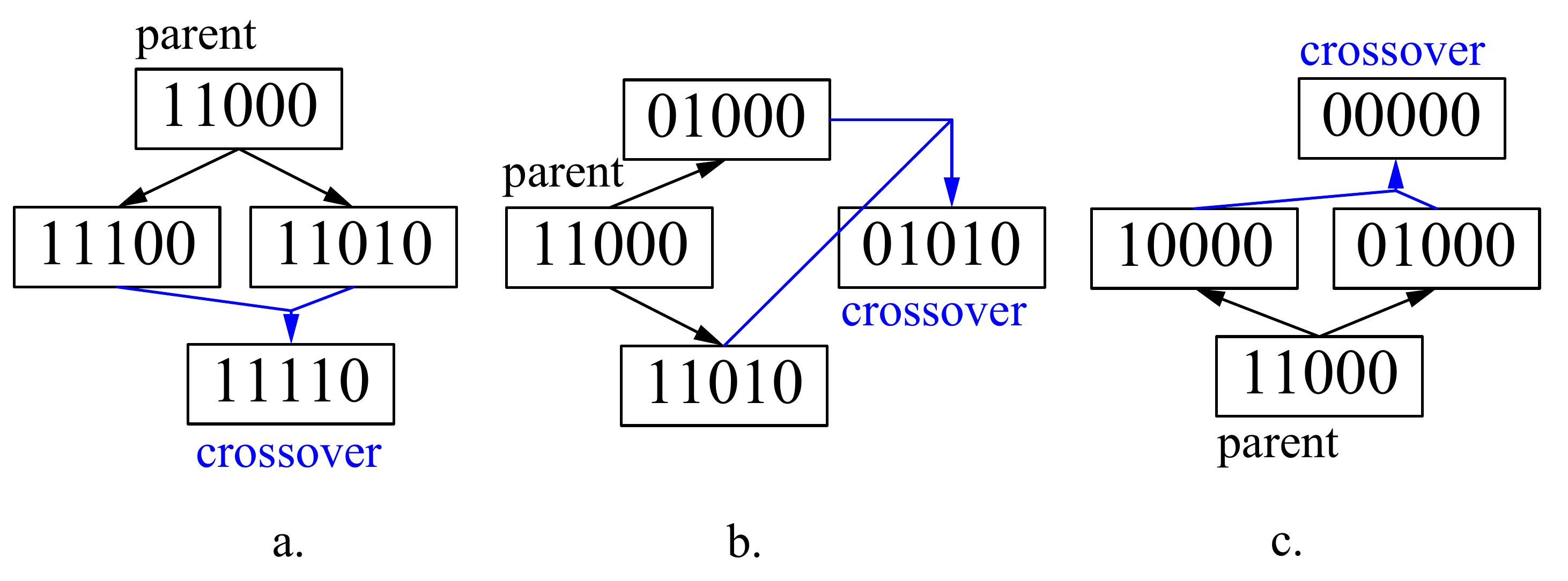}
\caption{3 types of cross over operations over the best and second best children of one parent. a. merge features from both children, equivalent to skip down; b. add one feature to one child, remove one feature from another child, equivalent to replace; c. remove one feature from each child, equivalent to skip up. All three operations can be represented by the simple arithmetic operation: $V_{crossover}$ = $V_{child1}$ + $V_{child2}$ - $V_{parent}$}
\label{graphjump}
\end{figure}

The idea of crossover operator comes from the genetic algorithm \cite{yang1998feature}, a classical meta-heuristic optimization approach that simulates natural selection process. The core of the genetic algorithm is mutation and crossover operator. Mutation randomly changes one or several bits of population. Crossover takes the two best vertices that share the same parent and generates a new child from 3 possible operations as illustrated in Fig.~\ref{graphjump}.

After a node with the best score in the current queue is popped out, the BFS with crossover adds all its children to the priority queue. While this step is the same as Greedy BFS, a crossover operation is conducted between the best two children (``first'' and ``second'' in Algorithm~\ref{alg:astarcomp}) of the node to identify a crossover node. This node is also added to the queue. There are three possible conditions depending on the relation of the two children with their parent (See Fig.~\ref{graphjump}).

\begin{table*}[h]
\caption{Prediction performance using gradient boosting tree and different feature selection method. For the BFS method, the patience parameter is set to 25. For GBT, number of tree = 100, with depth searched from 2 to 5. Columns 2-6 are $r^2$ scores.}
\label{tab:compare}
\centering
\begin{tabular}{|c|c|c|c|c|c|c|c|}
    \hline
       & Meta-heuristic & \multicolumn{2}{|c|}{Greedy} & \multicolumn{2}{|c|}{Best First Search} & \multicolumn{2}{|c|}{BFS with Crossover Results}  \\
    \hline
    Test/Cohort & Genetic Algorithm & Forward & Backward & Greedy BFS & BFS with Crossover & Pearson Coefficient & p-value\\ \hline
    DSF NC     & 0.4298 & 0.2074 & -0.0649 & 0.4529 & \textbf{0.5055}  & 0.75* & 0.0109\\ \hline
    DSB NC     & 0.4982 & 0.6200 & 0.2307 & \textbf{0.6408} & \textbf{0.6408} & 0.83* & 0.0051 \\ \hline
    LNS NC    & 0.3599 & 0.4206 & -0.1582 & 0.5182 & \textbf{0.5806} & 0.79* & 0.0138\\ \hline
    DSF mTBI   & 0.3510 & 0.3639 & -0.3072 & 0.4396 & \textbf{0.5090} & 0.74** & 0.0027\\ \hline
    DSB mTBI    & 0.5186 & 0.5080 & 0.1321 & 0.5193 & \textbf{0.6005} & 0.80** & 0.0005\\ \hline
    LNS mTBI    & 0.5370 & 0.4749 & -0.0763 & 0.5671 & \textbf{0.6036} & 0.82* & 0.0013\\ \hline
    DSF combine   & 0.2086 & 0.2895 & -0.1444 & 0.3709 & \textbf{0.3848} & 0.64** & 0.0019 \\ \hline
    DSB combine    & 0.2007 & 0.2089 & -0.1592 & 0.4075 & \textbf{0.4491} & 0.69** & 0.0018\\ \hline
    LNS combine   & 0.2055 & 0.3931 & 0.1702 & 0.4813 & \textbf{0.4874} & 0.72*** & 0.0003\\ \hline
\end{tabular}
\end{table*}

Compared to Greedy BFS, the crossover node, which is likely a good node with high score, will be evaluated with the same priority as all the children of the current node. With Greedy BFS, the crossover node will have to be evaluated along with all the children of the ``first'' node. With crossover, if the ``cross'' node is actually better than ``first'', the evaluations of other children of ``first'' will be skipped. However, there is no guarantee that the ``cross'' node is better than the children node of ``first'', so BFS with crossover may not always yield better results than Greedy BFS.

\subsection{Gradient Boosting Tree}

Gradient boosting tree (GBT) is chosen as an estimator for its simplicity and robustness. The idea of boosting is to combine the output of many weak models to produce a powerful ensemble \cite{hastie2013elements}. Gradient boosting adds the idea of steepest descent on top of boosting \cite{friedman2001greedy}. It iteratively adds new weak model to correct the largest previous error. In addition, decision tree is often chosen as a weak estimator. GBT has strong generalization ability and robustness to errors \cite{hastie2013elements}. In our preliminary work, we have compared GBT with other regression methods including Support Vector Machine and Neural Network. For the $r^2$ result of Greedy DSB NC test, SVM and NN achieve $0.3455$ and $0.2501$ respectively, which is significantly lower than GBT ($0.6200$). Hence, we choose to present only the performance of GBT under different feature selection methods.

\subsection{Repeated Stratified K-fold Cross Validation}

K-fold Cross-validation (CV) is widely applied method to estimate model performance \cite{hastie2013elements}. Here, we use 5-fold cross validation with stratified CV split, which splits the entire dataset into 5 folds with the same distribution of labels \cite{zhang2016facial}\cite{hastie2013elements}. In our case the labels are continuously distributed in $[-3, 3]$. We quantize this range to 5 bins and each fold would have the same percentage of the samples in each bin, as the whole dataset. 

\section{Results and Discussion}

\subsection{Prediction results between BFS with Crossover and other heuristic algorithms}

For each NP test, we develop three GBT models using control subjects, mTBI subjects, and all subjects, respectively. For each model, we perform feature selection using the proposed BFS with Crossover method as well as several other methods including greedy forward, greedy backward and genetic algorithm.

The average $r^2$ score among all validation samples is chosen to assess of model performance. $r^2$ is defined as the portion of variance explained: 
\begin{equation}\label{r2}
r^2 \gets 1 - \frac{MSE(y, \hat{y})}{Var(y)}
\end{equation}

\begin{figure}[ht]
\centering
\includegraphics[width = \linewidth]{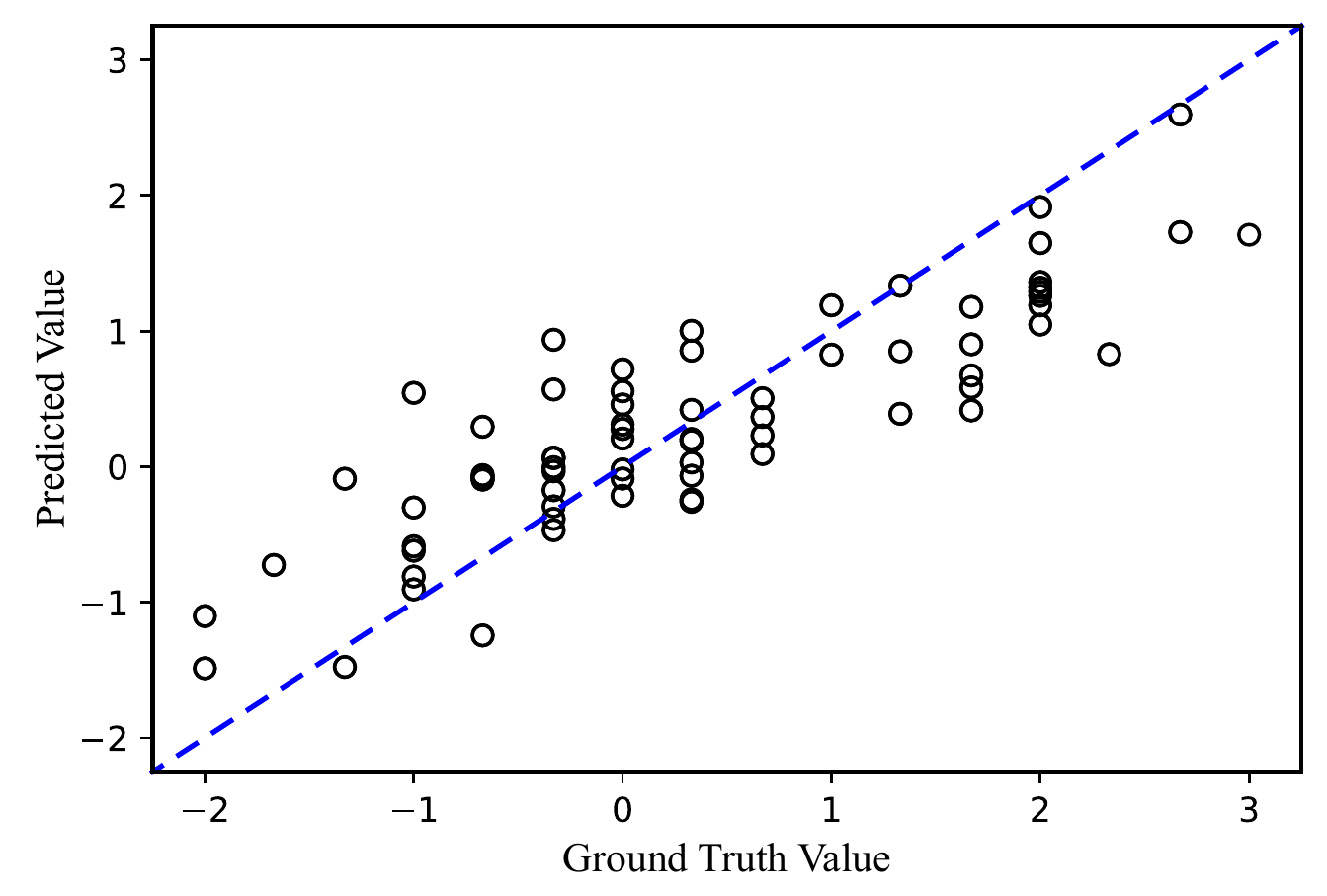}
\caption{Comparison between ground truth label values and predicted label values for the DSB test, using the model developed for the mTBI population. Data shown here are from the validation samples in all five folds.}
\label{graphcp}
\end{figure}

Table \ref{tab:compare} summarizes the performance of different feature selection methods. Greedy backward search yields poor performance, suggesting that this is not a viable method when the feature space is very large. The greedy forward and genetic algorithm are substantially better than greedy backward, but the $r^2$ score is still mostly below 0.5 (See Tab.~\ref{tab:compare}). The Greedy BFS provides substantial improvement over these two methods in all the models. Finally, BFS with crossover achieves further improvement over greedy BFS in all cases, most of which yield $r^2$ scores above 0.5. Especially, the performance on DSB mTBI improves from 0.5193 by Greedy BFS to 0.6005 by BFS with crossover (See Tab.~\ref{tab:compare}). The relatively high $r^2$ scores and the scatter plot in Fig.~\ref{graphcp} indicate a reasonably good fit. 

Last two columns of Table \ref{tab:compare} present the Pearson correlation between the ground truth and the predicted values by BFS with crossover and the corresponding p-value, which indicates the probability that an uncorrelated system produces such computed Pearson correlation. It could be observed that for most of the tests the Pearson correlation is larger than 0.7 with p value less than 0.05, which for biological systems indicates a strong and reliable relationship.

\subsection{Selected Features}

\begin{table}[h]
\caption{Selected MRI Metrics by BFS with Crossover}
\label{tab:analysis}
\centering
\begin{tabularx}{\columnwidth}{|c|X|}
    \hline
    Test/Cohort  & Metrics Chosen Frequency\\
    \hline
    DSF NC      & FA(5), De-perp(4), De-par(4), MK(2), AK(2), AWF(1), DA(1), MD(0) \\
    \hline
    DSB NC      & De-par(3), De-perp(2), MK(2), FA(2), AK(2), MD(1), AWF(1), DA(0) \\
    \hline
    LNS NC  & DA(4), AWF(3), De-perp(2), De-par(1), AK(1), MK(0), MD(0), FA(0) \\
    \hline
    NC SUM     & De-perp(8), De-par(8), DA(5), AWF(5), AK(5), FA(5), MK(4), MD(1) \\
    \hline
    DSF mTBI     & De-perp(5), De-par(4), AWF(3), MK(2), DA(2), AK(1), MD(0), FA(0) \\
    \hline
    DSB mTBI  & AWF(10), AK(5), MK(3), FA(3), DA(2), MD(1), De-perp(1), De-par(0) \\
    \hline
    LNS mTBI     & DA(6), AWF(3), De-par(3), AK(3), MK(2), MD(1), FA(1), De-perp(1) \\
    \hline
    mTBI SUM     & AWF(16), DA(10), AK(9), De-perp(7), De-par(7), MK(7), FA(4), MD(2) \\
    \hline
    DSF combine     & MD(5), DA(4), De-par(4), AWF(2), AK(2), FA(1), MK(0), De-perp(0) \\
    \hline
    DSB combine      & AWF(4), DA(3), De-perp(3), MK(2), De-par(2), AK(2), MD(1), FA(1) \\
    \hline
    LNS combine    & FA(5), AK(5), De-par(4), AWF(4), DA(2), De-perp(2), MK(1), MD(1) \\
    \hline
    combine SUM     & AWF(16), DA(10), De-par(10), AK(9), De-perp(7), FA(7), MK(3), MD(2) \\
    \hline
\end{tabularx}
\end{table}

The features chosen by BFS with crossover are analyzed since they produce best accuracy. The number of times a diffusion MRI metric is chosen is accumulated and summarized in Table~\ref{tab:analysis}.

For predicting the LNS test performance, it is interesting to observe that DA (Intra-axonal diffusivity, See Table~\ref{tab:metric}) metric is selected most often, for the modeled developed for the NC and mTBI cohorts, respectively. LNS is the most complex working memory task among these three tests and may have greater dependency on specific microstructural integrity more so than easy tasks. DA reflects axon injury or integrity and has been previously implicated in mTBI  \cite{chung2018white}.

Additionally, it is noted that when counting the total
number of times a metric is chosen over all three working
memory tests, we see that for the models developed for the
mTBI and control populations, respectively, the most frequently chosen features include De-par, De-perp, AWF, and DA. These compartment-specific metrics have been shown to be more sensitive to the underlying microstructure than others, such as DTI and DKI, which are known to be non-specific and empiric (See Table~\ref{tab:metric}).

Comparing the performances of separate models for the different cohorts (See Table~\ref{tab:compare}), we see that we are able to predict well with the models for the mTBI and NC cohorts, respectively. Furthermore, we see that the combined models (mTBI and NC together) are not as good with a weaker correlation coefficient for all three prediction tasks. The features chosen among these three populations for predicting the same NP score also differ (Table \ref{tab:analysis}). This suggests that mTBI and NP are two distinct populations in terms of white matter microstructure, in keeping with what we know about mTBI and white matter injury. 

\section{Conclusion}

In this work, a new feature selection algorithm for predicting performance on working memory using diffusion MRI features is proposed. The algorithm is able to search over a large feature space effectively and achieved consistently better performance than other popular feature selection methods. This novel feature selection method is applicable to other classification and regression problems with large feature space and limited training data.

The prediction models using the selected features achieved quite high Pearson Correlation ($>0.7$ in all cases) with very low p-value ($<0.002$), demonstrating statistically significant agreement between the predicted scores and the measured working memory test scores. These results suggest that optimizing feature selection for predicting NP test performance has a great potential to reveal the most important imaging features that would be related to cognitive functions or cognitive impairments in mTBI patients.

\section*{Acknowledgment}

Research reported in this paper is supported in part by grant funding from the National Institute for Neurological Disorders and Stroke (NINDS), National Institutes of Health (NIH): R21 NS090349, R01 NS039135-11, R01 NS088040 and NIBIB Biomedical Technology Resource Center Grant NIH P41 EB01718. The content is solely the responsibility of the authors and does not necessarily represent the official views of the NIH.

\bibliographystyle{IEEEtran}
\bibliography{mybibfile}

\begin{thebibliography}{10}
\providecommand{\url}[1]{#1}
\csname url@samestyle\endcsname
\providecommand{\newblock}{\relax}
\providecommand{\bibinfo}[2]{#2}
\providecommand{\BIBentrySTDinterwordspacing}{\spaceskip=0pt\relax}
\providecommand{\BIBentryALTinterwordstretchfactor}{4}
\providecommand{\BIBentryALTinterwordspacing}{\spaceskip=\fontdimen2\font plus
\BIBentryALTinterwordstretchfactor\fontdimen3\font minus
  \fontdimen4\font\relax}
\providecommand{\BIBforeignlanguage}[2]{{%
\expandafter\ifx\csname l@#1\endcsname\relax
\typeout{** WARNING: IEEEtran.bst: No hyphenation pattern has been}%
\typeout{** loaded for the language `#1'. Using the pattern for}%
\typeout{** the default language instead.}%
\else
\language=\csname l@#1\endcsname
\fi
#2}}
\providecommand{\BIBdecl}{\relax}
\BIBdecl

\bibitem{faul2010traumatic}
M.~Faul, M.~M. Wald, L.~Xu, and V.~G. Coronado, ``Traumatic brain injury in the
  united states; emergency department visits, hospitalizations, and deaths,
  2002-2006,'' 2010.

\bibitem{voormolen2018divergent}
D.~C. Voormolen, M.~C. Cnossen, S.~Polinder, N.~Von~Steinbuechel, P.~E. Vos,
  and J.~A. Haagsma, ``Divergent classification methods of post-concussion
  syndrome after mild traumatic brain injury: prevalence rates, risk factors,
  and functional outcome,'' \emph{Journal of neurotrauma}, vol.~35, no.~11, pp.
  1233--1241, 2018.

\bibitem{grossman2010mild}
E.~J. Grossman, M.~Inglese, and R.~Bammer, ``Mild traumatic brain injury: is
  diffusion imaging ready for primetime in forensic medicine?'' \emph{Topics in
  magnetic resonance imaging: TMRI}, vol.~21, no.~6, p. 379, 2010.

\bibitem{shenton2012review}
M.~E. Shenton, H.~Hamoda, J.~Schneiderman, S.~Bouix, O.~Pasternak, Y.~Rathi,
  M.-A. Vu, M.~P. Purohit, K.~Helmer, I.~Koerte \emph{et~al.}, ``A review of
  magnetic resonance imaging and diffusion tensor imaging findings in mild
  traumatic brain injury,'' \emph{Brain imaging and behavior}, vol.~6, no.~2,
  pp. 137--192, 2012.

\bibitem{chung2018white}
S.~Chung, E.~Fieremans, X.~Wang, N.~E. Kucukboyaci, C.~J. Morton, J.~Babb,
  P.~Amorapanth, F.-Y.~A. Foo, D.~S. Novikov, S.~R. Flanagan \emph{et~al.},
  ``White matter tract integrity: an indicator of axonal pathology after mild
  traumatic brain injury,'' \emph{Journal of neurotrauma}, vol.~35, no.~8, pp.
  1015--1020, 2018.

\bibitem{inpresspaper}
S.~Chung, X.~Wang, E.~Fieremans, R.~Joseph, A.~Prin, F.~Farng-Yang~A,
  C.~Morton, N.~Dmitry, F.~Steven~R, and Y.~W. Lui, ``Altered relationship
  between working memory and brain microstructure after mild traumatic brain
  injury,'' \emph{American Journal of Neuroradiology}, in press.

\bibitem{miles2008short}
L.~Miles, R.~I. Grossman, G.~Johnson, J.~S. Babb, L.~Diller, and M.~Inglese,
  ``Short-term dti predictors of cognitive dysfunction in mild traumatic brain
  injury,'' \emph{Brain injury}, vol.~22, no.~2, pp. 115--122, 2008.

\bibitem{lui2014classification}
Y.~W. Lui, Y.~Xue, D.~Kenul, Y.~Ge, R.~I. Grossman, and Y.~Wang,
  ``Classification algorithms using multiple mri features in mild traumatic
  brain injury,'' \emph{Neurology}, vol.~83, no.~14, pp. 1235--1240, 2014.

\bibitem{RN1}
S.~Minaee, Y.~Wang, and Y.~W. Lui, ``Prediction of longterm outcome of
  neuropsychological tests of mtbi patients using imaging features,'' in
  \emph{2013 IEEE Signal Processing in Medicine and Biology Symposium
  (SPMB)}.\hskip 1em plus 0.5em minus 0.4em\relax IEEE, Conference Proceedings,
  pp. 1--6.

\bibitem{RN2}
S.~Minaee, Y.~Wang, A.~Aygar, S.~Chung, X.~Wang, Y.~W. Lui, E.~Fieremans,
  S.~Flanagan, and J.~Rath, ``Mtbi identification from diffusion mr images
  using bag of adversarial visual features,'' \emph{IEEE transactions on
  medical imaging}, 2019.

\bibitem{makhzani2015adversarial}
A.~Makhzani, J.~Shlens, N.~Jaitly, I.~Goodfellow, and B.~Frey, ``Adversarial
  autoencoders,'' \emph{arXiv preprint arXiv:1511.05644}, 2015.

\bibitem{kao2016unsupervised}
P.-Y. Kao, E.~Rojas, J.~W. Chen, A.~Zhang, and B.~Manjunath, ``Unsupervised 3-d
  feature learning for mild traumatic brain injury,'' in \emph{International
  Workshop on Brainlesion: Glioma, Multiple Sclerosis, Stroke and Traumatic
  Brain Injuries}.\hskip 1em plus 0.5em minus 0.4em\relax Springer, 2016, pp.
  282--290.

\bibitem{mitra2016statistical}
J.~Mitra, K.-k. Shen, S.~Ghose, P.~Bourgeat, J.~Fripp, O.~Salvado, K.~Pannek,
  D.~J. Taylor, J.~L. Mathias, and S.~Rose, ``Statistical machine learning to
  identify traumatic brain injury (tbi) from structural disconnections of white
  matter networks,'' \emph{NeuroImage}, vol. 129, pp. 247--259, 2016.

\bibitem{RN13}
R.~Kohavi and G.~H. John, ``Wrappers for feature subset selection,''
  \emph{Artificial intelligence}, vol.~97, no. 1-2, pp. 273--324, 1997.

\bibitem{sattler2009assessment}
J.~M. Sattler and J.~J. Ryan, \emph{Assessment with the WAIS-IV}.\hskip 1em
  plus 0.5em minus 0.4em\relax Jerome M Sattler Publisher, 2009.

\bibitem{fieremans2011white}
E.~Fieremans, J.~H. Jensen, and J.~A. Helpern, ``White matter characterization
  with diffusional kurtosis imaging,'' \emph{Neuroimage}, vol.~58, no.~1, pp.
  177--188, 2011.

\bibitem{jensen2017evaluating}
J.~H. Jensen, E.~T. McKinnon, G.~R. Glenn, and J.~A. Helpern, ``Evaluating
  kurtosis-based diffusion mri tissue models for white matter with fiber ball
  imaging,'' \emph{NMR in Biomedicine}, vol.~30, no.~5, p. e3689, 2017.

\bibitem{RN15}
G.~Chandrashekar and F.~Sahin, ``A survey on feature selection methods,''
  \emph{Computers and Electrical Engineering}, vol.~40, no.~1, pp. 16--28,
  2014.

\bibitem{RN18}
D.~Rodrigues, L.~A. Pereira, R.~Y. Nakamura, K.~A. Costa, X.-S. Yang, A.~N.
  Souza, and J.~P. Papa, ``A wrapper approach for feature selection based on
  bat algorithm and optimum-path forest,'' \emph{Expert Systems with
  Applications}, vol.~41, no.~5, pp. 2250--2258, 2014.

\bibitem{doran1966experiments}
J.~E. Doran and D.~Michie, ``Experiments with the graph traverser program,''
  \emph{Proceedings of the Royal Society of London. Series A. Mathematical and
  Physical Sciences}, vol. 294, no. 1437, pp. 235--259, 1966.

\bibitem{RN22}
H.~Arai, C.~Maung, and H.~Schweitzer, ``Optimal column subset selection by
  a-star search,'' in \emph{Twenty-ninth AAAI conference on artificial
  intelligence}, 2015.

\bibitem{pudil1994floating}
P.~Pudil, J.~Novovi{\v{c}}ov{\'a}, and J.~Kittler, ``Floating search methods in
  feature selection,'' \emph{Pattern recognition letters}, vol.~15, no.~11, pp.
  1119--1125, 1994.

\bibitem{yang1998feature}
J.~Yang and V.~Honavar, ``Feature subset selection using a genetic algorithm,''
  in \emph{Feature extraction, construction and selection}.\hskip 1em plus
  0.5em minus 0.4em\relax Springer, 1998, pp. 117--136.

\bibitem{hastie2013elements}
\BIBentryALTinterwordspacing
T.~Hastie, R.~Tibshirani, and J.~Friedman, \emph{The Elements of Statistical
  Learning: Data Mining, Inference, and Prediction}, ser. Springer Series in
  Statistics.\hskip 1em plus 0.5em minus 0.4em\relax Springer New York, 2013.
  [Online]. Available: \url{https://books.google.com/books?id=yPfZBwAAQBAJ}
\BIBentrySTDinterwordspacing

\bibitem{friedman2001greedy}
J.~H. Friedman, ``Greedy function approximation: a gradient boosting machine,''
  \emph{Annals of statistics}, pp. 1189--1232, 2001.

\bibitem{zhang2016facial}
Y.-D. Zhang, Z.-J. Yang, H.-M. Lu, X.-X. Zhou, P.~Phillips, Q.-M. Liu, and
  S.-H. Wang, ``Facial emotion recognition based on biorthogonal wavelet
  entropy, fuzzy support vector machine, and stratified cross validation,''
  \emph{IEEE Access}, vol.~4, pp. 8375--8385, 2016.

\end{thebibliography}


\begin{thebibliography}{10}

\bibitem{kuo2015automatic}
Jen-wei Kuo, Yao Wang, Orlando Aristizabal, Daniel~H Turnbull, Jeffrey
  Ketterling, and Jonathan Mamou,
\newblock ``Automatic mouse embryo brain ventricle segmentation, gestation
  stage estimation, and mutant detection from 3d 40-mhz ultrasound data,''
\newblock in {\em 2015 IEEE International Ultrasonics Symposium (IUS)}. IEEE,
  2015, pp. 1--4.

\bibitem{aristizabal2013high}
Orlando Aristiz{\'a}bal, Jonathan Mamou, Jeffrey~A Ketterling, and Daniel~H
  Turnbull,
\newblock ``High-throughput, high-frequency 3-d ultrasound for in utero
  analysis of embryonic mouse brain development,''
\newblock {\em Ultrasound in Medicine \& Biology}, vol. 39, no. 12, pp.
  2321--2332, 2013.

\bibitem{henkelman2010systems}
R~Mark Henkelman,
\newblock ``Systems biology through mouse imaging centers: experience and new
  directions,''
\newblock {\em Annual Review of Biomedical Engineering}, vol. 12, pp. 143--166,
  2010.

\bibitem{kuo2015nested}
Jen-wei Kuo, Jonathan Mamou, Orlando Aristiz{\'a}bal, Xuan Zhao, Jeffrey~A
  Ketterling, and Yao Wang,
\newblock ``Nested graph cut for automatic segmentation of high-frequency
  ultrasound images of the mouse embryo,''
\newblock {\em IEEE Transactions on Medical Imaging (TMI)}, vol. 35, no. 2, pp.
  427--441, 2015.

\bibitem{kuo2018automatic}
Jen-wei Kuo, Ziming Qiu, Orlando Aristizabal, Jonathan Mamou, Daniel~H
  Turnbull, Jeffrey Ketterling, and Yao Wang,
\newblock ``Automatic body localization and brain ventricle segmentation in 3d
  high frequency ultrasound images of mouse embryos,''
\newblock in {\em 2018 IEEE 15th International Symposium on Biomedical Imaging
  (ISBI)}. IEEE, 2018, pp. 635--639.

\bibitem{long2015fully}
Jonathan Long, Evan Shelhamer, and Trevor Darrell,
\newblock ``Fully convolutional networks for semantic segmentation,''
\newblock in {\em Proceedings of the IEEE Conference on Computer Vision and
  Pattern Recognition (CVPR)}, 2015, pp. 3431--3440.

\bibitem{qiu2018deep}
Ziming Qiu, Jack Langerman, Nitin Nair, Orlando Aristizabal, Jonathan Mamou,
  Daniel~H Turnbull, Jeffrey Ketterling, and Yao Wang,
\newblock ``Deep bv: A fully automated system for brain ventricle localization
  and segmentation in 3d ultrasound images of embryonic mice,''
\newblock in {\em 2018 IEEE Signal Processing in Medicine and Biology Symposium
  (SPMB)}. IEEE, 2018, pp. 1--6.

\bibitem{qiu2019automatic}
Ziming Qiu, Nitin Nair, Jack Langerman, Orlando Aristizabal, Jonathan Mamou,
  Daniel~H Turnbull, Jeffrey~A Ketterling, and Yao Wang,
\newblock ``Automatic mouse embryo brain ventricle \& body segmentation and
  mutant classification from ultrasound data using deep learning,''
\newblock {\em arXiv preprint arXiv:1909.10555}, 2019.

\bibitem{roth2018application}
Holger~R Roth, Hirohisa Oda, Xiangrong Zhou, Natsuki Shimizu, Ying Yang,
  Yuichiro Hayashi, Masahiro Oda, Michitaka Fujiwara, Kazunari Misawa, and
  Kensaku Mori,
\newblock ``An application of cascaded 3d fully convolutional networks for
  medical image segmentation,''
\newblock {\em Computerized Medical Imaging and Graphics}, vol. 66, pp. 90--99,
  2018.

\bibitem{tang2019multi}
Yujiao Tang, Feng Yang, Shaofeng Yuan, et~al.,
\newblock ``A multi-stage framework with context information fusion structure
  for skin lesion segmentation,''
\newblock in {\em 2019 IEEE 16th International Symposium on Biomedical Imaging
  (ISBI)}. IEEE, 2019, pp. 1407--1410.

\bibitem{tu2008auto}
Zhuowen Tu,
\newblock ``Auto-context and its application to high-level vision tasks,''
\newblock in {\em 2008 IEEE Conference on Computer Vision and Pattern
  Recognition (CVPR)}. IEEE, 2008, pp. 1--8.

\bibitem{milletari2016v}
Fausto Milletari, Nassir Navab, and Seyed-Ahmad Ahmadi,
\newblock ``V-net: Fully convolutional neural networks for volumetric medical
  image segmentation,''
\newblock in {\em 2016 Fourth International Conference on 3D Vision (3DV)}.
  IEEE, 2016, pp. 565--571.

\bibitem{paszke2017automatic}
Adam Paszke, Sam Gross, Soumith Chintala, Gregory Chanan, Edward Yang, Zachary
  DeVito, Zeming Lin, Alban Desmaison, Luca Antiga, and Adam Lerer,
\newblock ``Automatic differentiation in pytorch,''
\newblock in {\em NIPS}, 2017.

\bibitem{kingma2014adam}
Diederik~P Kingma and Jimmy Ba,
\newblock ``Adam: A method for stochastic optimization,''
\newblock {\em arXiv preprint arXiv:1412.6980}, 2014.

\end{thebibliography}

\end{document}